# DYNAMICS OF INTENSE PARTICLE BEAM IN AXIAL-SYMMETRIC MAGNETIC FIELD

Yuri K. Batygin

Los Alamos National Laboratory, Los Alamos, NM 87545, USA

*Abstract*

Axial-symmetric magnetic field is often used in focusing of particle beams. Most existing ion Low Energy Beam Transport lines are based on solenoid focusing. Modern accelerator projects utilize superconducting solenoids in combination with superconducting accelerating cavities for acceleration of high-intensity particle beams. Present paper discusses conditions for matched beam in axial-symmetric magnetic field. Analysis allows us to minimize power consumption of solenoids and beam emittance growth due to nonlinear space charge, lens aberrations, and maximize acceptance of the channel. Expressions for maximum beam current in focusing structure, beam emittance growth due to spherical aberrations and non-linear space charge forces are derived.

## 1. LATTICE OF PERIODIC SOLENOID CHANNEL

Consider a focusing lattice consisting of a periodic sequence of focusing solenoids of length $D$, field $B_o$, distance between lenses $l$, and period $L = l + D$ (see Fig. 1). A matched beam reaches its maximum size in the center of the solenoids, and minimum size in the middle of drift space (see Fig. 2). The transformation matrix in a rotating frame through a period of the structure between centers of solenoids is given by [1]

$$\begin{pmatrix} \cos\frac{\theta}{2} & \frac{D}{\theta}\sin\frac{\theta}{2} \\ -\frac{\theta}{D}\sin\frac{\theta}{2} & \cos\frac{\theta}{2} \end{pmatrix} \begin{pmatrix} 1 & l \\ 0 & 1 \end{pmatrix} \begin{pmatrix} \cos\frac{\theta}{2} & \frac{D}{\theta}\sin\frac{\theta}{2} \\ -\frac{\theta}{D}\sin\frac{\theta}{2} & \cos\frac{\theta}{2} \end{pmatrix}$$

$$= \begin{pmatrix} \cos\theta - \frac{l}{2D}\theta\sin\theta & \frac{D}{\theta}\sin\theta + l\cos^2\frac{\theta}{2} \\ -\frac{\theta}{D}\sin\theta + l(\frac{\theta}{D})^2 \sin^2\frac{\theta}{2} & \cos\theta - \frac{l}{2D}\theta\sin\theta \end{pmatrix}, \quad (1.1)$$

where $\theta$ is the rotational angle of particle trajectory in a solenoid:

$$\theta = \frac{qB_o D}{2mc\beta\gamma} \,. \quad (1.2)$$

The matrix of transformation through the period of the structure between centers of drift space is:

$$\begin{pmatrix} 1 & \frac{l}{2} \\ 0 & 1 \end{pmatrix} \begin{pmatrix} \cos\theta & \frac{D}{\theta}\sin\theta \\ -\frac{\theta}{D}\sin\theta & \cos\theta \end{pmatrix} \begin{pmatrix} 1 & \frac{l}{2} \\ 0 & 1 \end{pmatrix}$$

$$= \begin{pmatrix} \cos\theta - \frac{l}{2D}\theta\sin\theta & l\cos\theta - \frac{l^2\theta}{4D}\sin\theta + D\frac{\sin\theta}{\theta} \\ -\frac{\theta}{D}\sin\theta & \cos\theta - \frac{l}{2D}\theta\sin\theta \end{pmatrix}. \quad (1.3)$$

From the matrices, Eqs. (1.1), (1.3), the value of betatron tune shift per period, $\mu_o$, is determined by

$$\cos\mu_o = \cos\theta - \theta\sin\theta \frac{(L-D)}{2D} \quad . \quad (1.4)$$

Adopting the expansions $\cos\xi = 1 - \xi^2/2 + \xi^4/24$ and $\sin\xi = \xi - \xi^3/6$, the value of betatron tune shift per period reads:

$$\mu_o = \theta\sqrt{\frac{L}{D}}\sqrt{1 - \frac{\theta^2}{6}[1 - \frac{1}{2}(\frac{D}{L} + \frac{L}{D})]} \quad . \quad (1.5)$$

Thus, the maximum and minimum values of the beta-function $\beta_{max/min} = m_{12}/\sin\mu_o$ in the channel are given by:

$$\beta_{max} = \frac{L\cos^2\frac{\theta}{2}[1 - \frac{D}{L}(1 - \frac{\tan\theta/2}{(\theta/2)})]}{\sin\mu_o} \quad , \quad (1.6)$$

$$\beta_{min} = \frac{(L-D)\cos\theta - \frac{(L-D)^2\theta}{4D}\sin\theta + D\frac{\sin\theta}{\theta}}{\sin\mu_o} \quad . \quad (1.7)$$

Eqs. (1.6), (1.7) determine the maximum $R_{max} = \sqrt{\beta_{max} \ni}$ and minimum $R_{min} = \sqrt{\beta_{min} \ni}$ matched envelope of the beam with unnormalized emittance, $\ni$, and negligible beam current, $I = 0$. Acceptance of the channel with aperture radius, $a$, is given by $A = a^2/\beta_{max}$:

$$A = \frac{a^2 \sin\mu_o}{L\cos^2\frac{\theta}{2}[1 - \frac{D}{L}(1 - \frac{\tan\theta/2}{(\theta/2)})]}. \quad (1.8)$$

The acceptance, Eq. (1.8), is maximized at a betatron tune shift within the range $0 < \mu_o < 180^o$ (see Fig. 3b, solid line).

## 2. THIN LENS APPROXIMATION

If the thickness of the lens is significantly smaller than the period of the structure, $D/L \ll 1$, focusing properties of the solenoid can be represented by the thin lens matrix with focal length $f$:

$$M = \begin{pmatrix} 1 & 0 \\ -\dfrac{1}{f} & 1 \end{pmatrix}, \qquad f = \dfrac{D}{\theta^2}. \tag{2.1}$$

Consequently, the betatron tune shift per period of the structure is determined from Eq. (1.4) as:

$$\cos\mu_o = 1 - \theta^2 \dfrac{L}{2D} = 1 - \dfrac{L}{2f}, \tag{2.2}$$

from which the value of $\mu_o$ is

$$\mu_o \approx \theta\sqrt{\dfrac{L}{D}} = \sqrt{\dfrac{L}{f}}. \tag{2.3}$$

From the condition $|\cos\mu_o| \leq 1$, the stability criteria for particle oscillations is expressed as $0 \leq L \leq 4f$ [2]. Accordingly, the acceptance of the channel is simplified from Eq. (1.8) as

$$A \approx \dfrac{a^2}{L}\sin\mu_o, \tag{2.4}$$

and has a maximum at the value of $\mu_o \approx \pi/2$. In this case, $\cos\mu_o = 1 - L/2f = 0$, and $f = L/2$ which expresses a condition of symmetry of the channel.

The values of beta-function from Eqs. (1.6), (1.7) are approximated as

$$\beta_{min} = \dfrac{L}{\sin\mu_o}(1 - \dfrac{\mu_o^2}{4}), \qquad \beta_{max} = \dfrac{L}{\sin\mu_o}. \tag{2.5}$$

From Eq. (2.2), the value of $\sin\mu_o = \sqrt{1 - \cos^2\mu_o}$ can be expressed as

$$\sin\mu_o = \sqrt{\dfrac{L}{f}(1 - \dfrac{L}{4f})}. \tag{2.6}$$

Correspondingly, the values of maximum $R_{max} = \sqrt{\beta_{max}\,\vartheta}$ and minimum $R_{min} = \sqrt{\beta_{min}\,\vartheta}$ matched beam sizes are given by [2]

$$R_{max} = \sqrt{\vartheta L}\,[\dfrac{4(\dfrac{f}{L})^2}{4(\dfrac{f}{L}) - 1}]^{1/4}, \qquad R_{min} = R_{max}\sqrt{1 - \dfrac{L}{4f}}. \tag{2.7}$$

Under the maximum acceptance condition, $L = 2f$, the matched beam sizes, Eqs. (2.7), evaluate to

$$R_{min} = \sqrt{\frac{\ni L}{2}}, \qquad R_{max} = \sqrt{\ni L}, \qquad (2.8)$$

from which the ratio of the beam sizes in a matched beam is $R_{max} / R_{min} = \sqrt{2}$.

### 3. DYNAMICS OF SPACE-CHARGE-DOMINATED BEAM

Analysis presented above gives us the ability to match a beam with negligible current. For analysis with non-zero current, $I \neq 0$, let us use the KV envelope equation for a round beam envelope $R(z)$ in an axially-symmetric channel [2]:

$$\frac{d^2R}{dt^2} + \omega_L^2 R - \frac{(\ni \beta c)^2}{R^3} - \frac{2Ic^2}{I_c R \beta \gamma^3} = 0, \qquad (3.1)$$

where $\omega_L(z) = qB(z)/(2m\gamma)$ is the Larmor oscillation frequency in magnetic field, and $I_c = 4\pi\varepsilon_o mc^3/q$ = $3.13 \times 10^7 A/Z$ [Amp] is the characteristic beam current. The magnetic field along the structure can be expanded in Fourier series:

$$B^2(z) = B_o^2 [\frac{D}{L} + \frac{2}{\pi} \sum_{n=1}^{\infty} \frac{1}{n} \sin(\frac{\pi n D}{L}) \cos(\frac{2\pi n z}{L})]. \qquad (3.2)$$

Substitution of expansion, Eq. (3.2), into envelope equation (3.1) gives:

$$\frac{d^2R}{dt^2} = -\frac{R}{2\pi}(\frac{qB_o}{m\gamma})^2 \sum_{n=1}^{\infty} \frac{1}{n} \sin(\frac{\pi n D}{L}) \cos(\frac{2\pi n \beta c t}{L}) - \frac{RD}{4L}(\frac{qB_o}{m\gamma})^2 + \frac{(\ni \beta c)^2}{R^3} + \frac{2Ic^2}{I_c R \beta \gamma^3}. \qquad (3.3)$$

Note that the envelope equation, Eq. (3.3), describes the oscillatory motion in a combination of two oscillating fields:

$$\ddot{R} = \sum_{n=1}^{\infty} F_n(R) \cos(\omega_n t) - \frac{\partial U(R)}{\partial R}, \qquad (3.4)$$

where the first term describes oscillatory field due to variation of magnetic field $B(z)$ along the channel

$$F_n(R) = -\frac{R}{2\pi n}(\frac{qB_o}{m\gamma})^2 \sin(\frac{\pi n D}{L}), \qquad \omega_n = \frac{2\pi n \beta c}{L}, \qquad (3.5)$$

and the second term determines oscillating function $U(R)$, which depends on variable beam radius $R$:

$$\frac{\partial U(R)}{\partial R} = \frac{RD}{4L}(\frac{qB_o}{m\gamma})^2 - \frac{(\ni \beta c)^2}{R^3} - \frac{2Ic^2}{I_c R \beta \gamma^3}. \qquad (3.6)$$

However, for $\mu_o \ll 2\pi$ variation of beam radius at the period of the structure is small, and function $U(R)$ can be approximately assumed to be $z$ - independent. Under these assumptions, according to the averaging method [3], envelope oscillations can be decomposed into a combination of slow variable $R_{aver}(t)$ and a small-amplitude, rapidly oscillating component $\xi(t)$:

$$R(t) = R_{aver}(t) + \xi(t) \ . \tag{3.7}$$

The dynamics of the slow variable is governed by the Hamiltonian

$$H = \frac{(dR_{aver}/dt)^2}{2} + \frac{1}{4}\sum_{n=1}^{\infty}\frac{F_n^2(R_{aver})}{\omega_n^2} + U(R_{aver}) \ , \tag{3.8}$$

while the oscillatory correction is determined entirely by the properties of the rapidly varying part of the field:

$$\xi(t) = -\sum_{n=1}^{\infty}\frac{F_n(R_{aver})}{\omega_n^2}\cos\omega_n t \ . \tag{3.9}$$

The equation of motion for the slow variable $R_{aver}(t)$ is derived from the Hamiltonian, Eq. (3.8):

$$\ddot{R}_{aver}(t) = -\frac{1}{2}\sum_{n=1}^{\infty}\frac{F_n}{\omega_n^2}\frac{dF_n}{dR_{aver}} - \frac{\partial U}{\partial R_{aver}} \ . \tag{3.10}$$

Substitution of Eqs. (3.5), (3.6) ino Eq. (3.10) gives

$$\ddot{R}_{aver} = -R_{aver}\Omega_r^2 + \frac{(\partial \beta c)^2}{R_{aver}^3} + \frac{2Ic^2}{I_c R_{aver}\beta\gamma^3} \ , \tag{3.11}$$

where the transverse frequency of single particle oscillations is

$$\Omega_r^2 = \frac{(\theta\beta c)^2}{LD} + \frac{1}{2}(\frac{\theta^2 L\beta c}{\pi^2 D^2})^2\sum_{n=1}^{\infty}\frac{\sin^2(\pi n\frac{D}{L})}{n^4} \ . \tag{3.12}$$

The sum in Eq. (3.12) can be rewritten as

$$\sum_{n=1}^{\infty}\frac{\sin^2(n\zeta)}{n^4} = \frac{1}{2}\sum_{n=1}^{\infty}\frac{1}{n^4} - \frac{1}{2}\sum_{n=1}^{\infty}\frac{\cos 2n\zeta}{n^4} \ , \tag{3.13}$$

where $\zeta = \pi D/L$. Calculation of separate sums reads [4]:

$$\sum_{n=1}^{\infty}\frac{1}{n^4} = \frac{\pi^4}{90} \ , \qquad \sum_{n=1}^{\infty}\frac{\cos n\xi}{n^4} = \frac{\pi^4}{90} - \frac{\pi^2\xi^2}{12} + \frac{\pi\xi^3}{12} - \frac{\xi^4}{48} \ . \tag{3.14}$$

Therefore, Eq. (3.13) reads:

$$\sum_{n=1}^{\infty}\frac{\sin^2(\pi n\frac{D}{L})}{n^4} = \frac{\pi^4}{3}(\frac{D}{L})^2[\frac{1}{2} - \frac{D}{L} + \frac{1}{2}(\frac{D}{L})^2] \ . \tag{3.15}$$

Consequently, substitution of Eq. (3.15) into Eq. (3.12) yield the transverse oscillation frequency:

$$\Omega_r^2 = (\frac{\beta c}{L})^2 [\theta^2 \frac{L}{D} + \frac{\theta^4}{12}(\frac{L}{D})^2 - \frac{\theta^4}{6}\frac{L}{D} + \frac{\theta^4}{12}] \; . \tag{3.16}$$

The expression in the square brackets in Eq. (3.16) is a square of the betatron tune shift, Eq. (1.5), which is related to the transverse oscillation frequency by

$$\mu_o = \Omega_r \frac{L}{\beta c} \; . \tag{3.17}$$

Therefore, the averaging method gives the same value for the betatron tune shift as the matrix method. Upon changing the independent variable in Eq. (3.11) from $t$ to $z$, the equation for the slow envelope variable becomes

$$\frac{d^2 R_{aver}}{dz^2} - \frac{\ni^2}{R_{aver}^3} + \frac{\mu_o^2}{L^2} R_{aver} - \frac{P^2}{R_{aver}} = 0 \; , \tag{3.18}$$

where $P^2$ is the generalized beam perveance:

$$P^2 = \frac{2I}{I_c \beta^3 \gamma^3} \; . \tag{3.19}$$

Retaining only the leading-order term in Eq (3.9), $\xi(z) \approx -F_1(R_{aver})\cos\omega_1 t / \omega_1^2$, the rapidly oscillating component of the beam envelope is given by:

$$\xi(z) = \frac{R_{aver}(z)}{2\pi^3}(\theta \frac{L}{D})^2 \sin(\pi \frac{D}{L})\cos(2\pi \frac{z}{L}) \; . \tag{3.20}$$

Finally, the solution of the envelope equation can be expressed as

$$R(z) = R_{aver}(z)(1 + \upsilon_{max} \cos 2\pi \frac{z}{L}) \; , \tag{3.21}$$

where the amplitude of envelope oscillations around average value $R_{aver}(z)$ is

$$\upsilon_{max} = \frac{\theta^2}{2\pi^3}(\frac{L}{D})^2 \sin(\pi \frac{D}{L}) \; . \tag{3.22}$$

The solution (3.21) describes the characteristics of a general unmatched beam, with slow variation of average radius $R_{aver}(z)$ superimposed on a fast oscillation with the period equal to the period of the structure. The relative amplitude of variation of beam envelope, $\upsilon_{max}$, Eq. (3.22), does not depend on either the beam current or the beam emittance [5]. Analysis of space-charge limited beam current in transport systems was performed in Refs. [5, 6, 7]. A matched beam corresponds to a constant value of the average beam envelope $R_{aver}(z) = \bar{R}_{aver}$ and can be determined from Eq. (3.18) by setting $R''_{aver}(z) = 0$ :

$$\bar{R}_{aver} = \bar{R}_{aver}(0) \sqrt{b_o + \sqrt{1+b_o^2}}, \qquad (3.23)$$

where $\bar{R}_{aver}(0)$ is the matched average beam size with negligible space charge,

$$\bar{R}_{aver}(0) = \sqrt{\frac{\ni L}{\mu_o}}, \qquad (3.24)$$

and $b_o$ is the space charge parameter:

$$b_o = \frac{1}{(\beta\gamma)^3} \frac{I}{I_c} (\frac{\bar{R}_{aver}(0)}{\ni})^2. \qquad (3.25)$$

The minimum and maximum matched beam envelope sizes in presence of space charge forces are given by:

$$R_{max/min} = \bar{R}_{aver}(1 \pm \upsilon_{max}). \qquad (3.26)$$

Maximum beam current is achieved when maximum beam size is set equal to the aperture of the channel $R_{max} = a$, which is determined from Eqs. (3.23) - (3.26) as

$$a = \sqrt{\frac{\ni L}{\mu_o}} \sqrt{b_o + \sqrt{1+b_o^2}} \, (1+\upsilon_{max}). \qquad (3.27)$$

For negligible beam intensity, $b_o = 0$, Eq. (3.27) describes the beam with maximum possible emittance (acceptance of the channel) approximated by envelope equation $\ni = A_{env}$:

$$a = \sqrt{\frac{A_{env} L}{\mu_o}} (1+\upsilon_{max}). \qquad (3.28)$$

Eq. (3.28) gives the following approximation to the acceptance of the channel (this equation should be compared to the exact expression for acceptance, Eq. (1.8):

$$A_{env} = \frac{a^2 \mu_o}{L (1+\upsilon_{max})^2}. \qquad (3.29)$$

From Eqs. (3.27) - (3.28) the maximum beam current is:

$$I_{max} = \frac{I_c}{2} \frac{\mu_o}{L} A_{env} (\beta\gamma)^3 [1 - (\frac{\ni}{A_{env}})^2]. \qquad (3.30)$$

Fig. 3 illustrates matched beam sizes, $R_{max/min}$ for a beam with negligible current, and acceptance of the channel, $A$, obtained from matrix analysis as well as that obtained utilizing the averaging approximation to the envelope equation, $A_{env}$. It is clear that validity of averaging method (smooth approximation) is limited by values of $\mu_o \leq 60^o$. Enhancement of the beam current leads to a decrease in the depressed betatron tune shift $\mu$ [5]

$$\mu = \mu_o (\sqrt{1+b_o^2} - b_o) \,, \tag{3.31}$$

and the smooth approximation progressively approaches the exact solution of the envelope equation (see Fig. 4).

Space-charge-induced beam emittance growth in a focusing channel is limited by the value (free energy effect [8]):

$$\frac{\ni_{eff}}{\ni} = \sqrt{1 + \frac{2}{(\beta\gamma)^3} \frac{I}{I_c} (\frac{\bar{R}_{aver}}{\ni})^2 \frac{\Delta W}{W_o}} \,, \tag{3.32}$$

where the free energy parameter $\Delta W / W_o$ depends on the beam distribution (see Table 1). From Eq. (3.32) it is clear that minimization of the beam radius in the channel is the way to reduce space-charge induced beam emittance growth [8, p. 315].

## 4. MAXIMUM SPACE-CHARGE LIMITED BEAM CURRENT

Eq. (3.30) gives an approximate value of the maximum beam current in a periodic structure. To determine a more exact value for space - charge limited beam current, consider the beam transport in drift space between lenses described by the envelope equation (3.1) without a focusing term:

$$\frac{d^2 R}{dz^2} - \frac{\ni^2}{R^3} - \frac{P^2}{R} = 0 \,. \tag{4.1}$$

Equation (4.1) can be integrated to yield [5]:

$$(\frac{dR}{dz})^2 = (\frac{dR}{dz})_o^2 + (\frac{\ni}{R_o})^2 (1 - \frac{R_o^2}{R^2}) + P^2 \ln(\frac{R}{R_o})^2 \,. \tag{4.2}$$

Now consider the space charge dominated regime, where the beam emittance can be neglected, $\ni \approx 0$. At the middle point between lenses, $z = z_o$, the beam has a waist size $R_o = R_{min}$ and zero divergence, $R_o^{'} = 0$. Thus, in this case, equation (4.2) can be rewritten as

$$(\frac{d\bar{R}}{dZ})^2 = \ln \bar{R} \,, \qquad \bar{R} = \frac{R}{R_{min}}, \qquad Z = \frac{\sqrt{2} Pz}{R_{min}} \,. \tag{4.3}$$

Consequently, expansion of the beam radius in drift space from $\bar{R} = 1$ to $\bar{R}_{max} = R_{max} / R_{min}$ is determined by the integral:

$$\frac{1}{\bar{R}_{max}} \int_1^{\bar{R}_{max}} \frac{d\bar{R}}{\sqrt{\ln \bar{R}}} = \sqrt{2} P \frac{(z - z_o)}{R_{max}} \,. \tag{4.4}$$

The left hand side of Eq. (4.4) has a maximum value of 1.082 for $\bar{R}_{max} = 2.35$ [9]. As already alluded to above, the maximum radius is achieved in the channel at the distance equal to half of the spacing between lenses, $z - z_o = L/2$, which in turn yields

$$\frac{P_{max} L}{\sqrt{2} R_{max}} = 1.082 . \quad (4.5)$$

From this expression, the maximum transported current in the channel is

$$I_{lim} = 1.17 I_c (\beta\gamma)^3 (\frac{R_{max}}{L})^2 . \quad (4.6)$$

The divergence of the beam at the lens can be estimated from Eq. (4.2):

$$\frac{dR_{max}}{dz} = \sqrt{\frac{4 I_{lim}}{I_c (\beta\gamma)^3} \ln(\frac{R_{max}}{R_{min}})} \approx 2 \frac{R_{max}}{L} . \quad (4.7)$$

Total change in the slope of the beam envelope at the lens has to be equal to twice the value of $dR_{max}/dz$ determined by Eq. (4.7). Therefore, the required focal length of the lenses is $f \approx L/4$, and the maximum space charge limited beam current is achieved in a structure where $\mu_o \approx 180^o$. Such transports are usually unstable, and can be used only with a limited number of focusing elements.

From Eq. (3.32) one can estimate space-charge induced beam emittance growth in this case. Assuming initial beam emitance is negligible, $э \approx 0$, and taking into account that average radius of matched beam is

$$\bar{R}_{aver} = \frac{1}{2}(R_{max} + \frac{R_{max}}{2.35}) = 0.712 R_{max} , \quad (4.8)$$

the space-charge induced beam emittance for beam transport with limited beam current, Eq. (4.6), is

$$э_{eff} = 1.09 \frac{R_{max}^2}{L} \sqrt{\frac{\Delta W}{W_o}} . \quad (4.9)$$

Ratio of space charge term to emittance term in the envelope equation, Eq. (4.1) gives an estimation of dominance of space charge forces over emitance in beam dynamics:

$$b = \frac{2}{(\beta\gamma)^3} \frac{I_{lim}}{I_c} (\frac{\bar{R}_{aver}}{э_{eff}})^2 = \frac{1}{(\frac{\Delta W}{W_o})} . \quad (4.10)$$

Because the value of free energy parameter is within the range of $\Delta W / W_o = 0....8 \cdot 10^{-2}$, the value of parameter $b$, Eq. (4.10), is larger than unit, $b \gg 1$, which indicates dominance of space charge in beam dynamics and allows us to use simplification of negligible emittance for the considered case.

## 5. APPLICATION TO INJECTOR LEBT

Most existing ion Low Energy Beam Transports (LEBT) utilize 2 or 3 solenoids with intermediate equipment (deflectors, bending magnets, Wien filters, emittance stations) to match the beam from the exit of ion source column to the subsequent RF structure. Consider a LEBT comprised of 2 solenoids, separated by a distance $L$ (see Fig. 5). The beam is characterized by a certain emittance $э$ and effective current $I = I_o(1-\eta)$, where $I_o$ is the total beam current and $\eta$ is the space-charge neutralization factor. Initial envelope parameters $R_s$, $R_s^{'}$ are determined by extraction conditions from the ion source column. Final beam parameters $R_f$, $R_f^{'}$ are determined by the matching conditions at the front end of the RF accelerator. The purpose of the analysis is then to find appropriate solenoid parameters, and distances $d_1$, $d_2$.

Analysis of the previous sections allows us to select a beam envelope, corresponding to minimal values of the beam size at the center of solenoid, $R_{max}$. In turn, minimization of the beam size $R_{max}$ allows us to minimize solenoid power consumption, effect of spherical aberrations, beam losses, and space-charge induced beam emittance growth. A two-solenoid system can be regarded as a part of a periodic focusing structure. Consider a beam with negligible current, but with a finite value of beam emittance (emittance-dominated beam). Evolution of the beam radius $R$ along drift space $z$ between solenoids as a function of initial radius $R_o$ and slope of the envelope $R_o^{'}$ is obtained by integration of Eq. (4.2) assuming $I = 0$ [5]:

$$\frac{R}{R_o} = \sqrt{(1+\frac{R_o^{'}}{R_o}z)^2 + (\frac{э}{R_o^2})^2 z^2} \ . \tag{5.1}$$

From the symmetry point of view it is clear, that a matched beam has a minimum size, or waist, $R_{min} = R_o$ in the mid-point of the drift space between lenses, and maximum size $R_{max}$ inside focusing elements. At the waist point, $R_o^{'} = 0$. Therefore from Eq. (5.1)

$$R_{max}^2 = R_{min}^2 + (\frac{э L}{2 R_{min}})^2 \ . \tag{5.2}$$

Equating $\partial R_{max} / \partial R_{min} = 0$ determines the minimal value of $R_{max}$ as a function of beam emittance and distance between lenses

$$\frac{\partial R_{max}}{\partial R_{min}} = \frac{1}{\sqrt{R_{min}^2 + (\frac{э L}{2 R_{min}})^2}} [R_{min} - \frac{1}{R_{min}^3}(\frac{э L}{2})^2] = 0 \ . \tag{5.3}$$

The solution of Eq. (5.3) is

$$R_{min}(0) = \sqrt{\frac{э L}{2}} \ , \qquad\qquad R_{max}(0) = \sqrt{э L} \ , \tag{5.4}$$

which coincides with Eqs. (2.8) for a periodic solution of a matched beam with zero current at a phase advance of $\mu_o \approx \pi/2$. Eq. (5.4) determines the minimum value of $R_{max}$ at given value of beam emittance and given distance between solenoids $L$.

Now consider the space-charge dominated regime, where beam emittance is negligible. Analysis presented in Section 4 determines the condition for transporting a beam with maximum current through drift space restricted by aperture $R_{max}$ and distance $L$. From Eq. (4.5)

$$R_{max} = \frac{L}{1.082}\sqrt{\frac{I}{I_c(\beta\gamma)^3}} \;, \qquad\qquad R_{min} = \frac{R_{max}}{2.35}. \tag{5.5}$$

Eqs. (5.5) determine conditions for minimizing the value of $R_{max}$ of the beam with given current $I$ in the structure limited by distance $L$. In a more general case, when both beam emittance and beam current are not negligible, the precise value of $R_{max}$ is determined by the variation of the value of $R_{min}$ at the mid-point between solenoids, $z = z_o$, and a subsequent search for the smallest beam size at the center of the solenoids $R_{max}$ via an exact solution of the envelope equation in drift space. Dominance of emittance or space charge on beam dynamics can be estimated through space charge parameter $b_o$, Eq. (3.25), where optimal value of average zero-current beam radius, Eq. (3.24), corresponds to $\mu_o \approx \pi/2$

$$\bar{R}_{aver}(0) = \sqrt{2\frac{\ni L}{\pi}}. \tag{5.6}$$

The value of $b_o \approx 0$ corresponds to the emittance-dominated regime, while values of $b_o > 1$ correspond to the space-charge dominated regime.

After determination of the minimal value of $R_{max}$, the distances $d_1$, $d_2$ are defined by integration of equation (4.2) to establish points where the beam radius evolves from initial value of $R_o$ to $R_{max}$ [5]:

$$z = \frac{R_o^2}{2\ni} \int_1^{(\frac{R_{max}}{R_o})^2} \frac{ds}{\sqrt{[1+(\frac{R_o R_o'}{\ni})^2]s + (\frac{PR_o}{\ni})^2 s\ln s - 1}}. \tag{5.7}$$

In Eq. (5.7), the values of $R_o$, $R_o'$ correspond to either $R_s$, $R_s'$ or $R_f$, $R_f'$. The slopes of beam envelopes at solenoids $R_1'$, $R_2'$ can be found from Eq. (4.2):

$$R' = \sqrt{(R_o')^2 + (\frac{\ni}{R_o})^2[1-(\frac{R_o}{R})^2] + \frac{2I}{I_c(\beta\gamma)^3}\ln(\frac{R}{R_o})^2}. \tag{5.8}$$

The values of $R_{1d}'$, $R_{2d}'$ are determined by Eq. (4.2) assuming $R_o = R_{min}$, $R_o' = 0$. Then, the focal lengths of the solenoids $f_1$, $f_2$, are determined by the total change in the slope of the beam at each solenoid:

$$f_1 = \frac{R_{max}}{|R_{1d}'|+|R_1'|}, \; f_2 = \frac{R_{max}}{|R_{2d}'|+|R_2'|}. \tag{5.9}$$

Subsequently, the magnetic field within each solenoid is determined from Eqs. (1.2), (2.1) as

$$B_o = \frac{2mc\beta\gamma}{q\sqrt{fD}}. \tag{5.10}$$

The described procedure allows us to perform an optimization of Low Energy Beam Transport from point of view beam size minimization and the reduction of effects associated with beam emittance growth.

# 6. SPHERICAL ABERRATION OF AXIAL-SYMMETRIC MAGNETIC LENS

Dynamics in an axial-symmetric field may be severely affected by spherical aberrations, which result in beam emittance growth. Spherical aberration coefficients are typically expressed through field integrals [10]. Analysis of the effect of spherical aberrations can be simplified by assuming particle radius does not change significantly inside the lens. Consider a single- particle equation of motion in a magnetic field

$$\ddot{r} + r(\frac{qB_z}{2m\gamma})^2 = 0 . \tag{6.1}$$

The magnetic field along the structure can be represented as

$$B_z(r,z) = B(z) - \frac{r^2}{4} B''(z) , \tag{6.2}$$

where $B(z)$ is the longitudinal component of magnetic field at the axis. Usually it can be approximated as [11]

$$B(z) = \frac{B_o}{1 + (\frac{z}{d})^n} , \tag{6.3}$$

where $B_o$ is the maximum value of magnetic field, and $d$ is the characteristic length. Parameter $n = 2$ corresponds to well-known Glazer model [12], for short length/diameter lenses, while in many cases the field of a solenoid is better approximated by $n = 4$, which provides a more flat distribution.

Let us integrate Eq. (6.1) assuming the particle radius remains constant during lens-crossing. The change in slope of particle trajectory follows from Eq. (6.1):

$$r' = r_o' - r \, (\frac{q}{2mc\beta\gamma})^2 \int_{-\infty}^{\infty} B_z^2 \, dz . \tag{6.4}$$

In the near-axis approximation, the longitudinal field component is a weak function of radius, therefore one can assume:

$$B_z^2 \approx B^2(z) - \frac{r^2}{2} B(z) B''(z) , \qquad B''(z) = \frac{nB_o z^{n-2}}{d^n} \frac{[(1-n) + (1+n)(\frac{z}{d})^n]}{[1 + (\frac{z}{d})^n]^3} . \tag{6.5}$$

Under this approximation, equation (6.4) can be rewritten in more common way:

$$r' = r_o' - \frac{r}{f}(1 + C_\alpha r^2) , \tag{6.6}$$

where the focal length $f$ and spherical aberration coefficient $C_\alpha$ are determined by the expressions

$$\frac{1}{f} = d\left(\frac{qB_o}{2mc\beta\gamma}\right)^2 \int_{-\infty}^{\infty} \frac{d\xi}{(1+\xi^n)^2}, \qquad C_\alpha = -\frac{1}{2} \frac{\int_{-\infty}^{\infty} B(z)B''(z)dz}{\int_{-\infty}^{\infty} B^2(z)dz}. \tag{6.7}$$

More specifically, the focal length is determined by Eq. (2.1) with the effective length of the solenoid, $D$:

$$D = d \int_{-\infty}^{\infty} \frac{d\xi}{(1+\xi^n)^2} = \begin{cases} \frac{\pi}{2}d \approx 1.57d, & n = 2 \\ \frac{3\pi}{4\sqrt{2}}d \approx 1.666d, & n = 4. \end{cases} \tag{6.8}$$

Calculation of the spherical aberration coefficient gives [11]:

$$C_\alpha = \frac{1}{d^2} \frac{\int_{-\infty}^{\infty} \frac{(1-3\xi^2)}{(1+\xi^2)^4} d\xi}{\int_{-\infty}^{\infty} \frac{d\xi}{(1+\xi^2)^2}} = \frac{1}{4d^2} \qquad n = 2, \tag{6.9}$$

$$C_\alpha = \frac{2}{d^2} \frac{\int_{-\infty}^{\infty} \frac{(3-5\xi^4)}{(1+\xi^4)^4} \xi^2 d\xi}{\int_{-\infty}^{\infty} \frac{d\xi}{(1+\xi^4)^2}} = \frac{5}{12d^2} \qquad n = 4. \tag{6.10}$$

The performed analysis corresponds to a weak focusing regime, when the radius of particle changes insignificantly while passing through the lens. In many applications, the spherical aberration coefficient can be expressed through solenoid sizes as [13]

$$C_\alpha = \frac{5}{(S+2a)^2}, \tag{6.11}$$

where $2a$ is the pole piece diameter and $S$ is the solenoid pole gap width.

## 7. EFFECT OF SPHERICAL ABERRATION ON BEAM EMITTANCE GROWTH

Let us estimate the emittance growth of a beam during it's passage through the lens. We assume that the position of the particle is not changed while crossing the lens, and only the slope of the particle trajectory is altered. The transformation of particle variables before $(r_o, r_o')$ to after $(r, r')$ lens-crossing is given by:

$$r = r_o, \qquad\qquad r' = r_o' - \frac{r}{f}(1+C_\alpha r^2). \tag{7.1}$$

Suppose, initial phase space volume is bounded by the ellipse

$$\frac{r_o^2}{R^2}э + \frac{r_o'^2}{э}R^2 = э.  \tag{7.2}$$

To find the deformation of the boundary of the beam phase space after passing through the lens, let us substitute the inverse transformation $r_o = r$, $r_o' = r' + (r/f)(1 + C_\alpha r^2)$ into the ellipse equation, Eq. (7.2). The boundary of the new phase space volume, occupied by the beam after passing through the lens at phase plane $(r, r')$ is given by:

$$\frac{r^2}{R^2}э + \frac{R^2}{э}(r' + \frac{r}{f} + C_\alpha \frac{r^3}{f})^2 = э.  \tag{7.3}$$

Let us introduce new variables $(T, \psi)$ arising from the transformation:

$$\frac{r}{R} = \sqrt{T}\cos\psi,  \tag{7.4}$$

$$(r' + \frac{r}{f})\frac{R}{э} = \sqrt{T}\sin\psi.  \tag{7.5}$$

In terms of new variables, the shape of the beam emittance after lens-crossing is

$$T + 2\upsilon T^2 \sin\psi \cos^3\psi + T^3 \upsilon^2 \cos^6\psi = 1,  \tag{7.6}$$

where the following notation is used

$$\upsilon = \frac{C_\alpha R^4}{f\,э}.  \tag{7.7}$$

Without nonlinear perturbations, $\upsilon = 0$ and Eq. (7.6) describes a circle in phase space. Conversely, if $\upsilon \neq 0$, Eq. (7.6) describes an $S$ – shape distorted figure of beam emittance (see Fig. 6). Filamentation of beam emittance in phase space is a fundamental property of a beam affected by aberrations.

Being symplectic in nature, the transformation, Eq. (7.1), conserves phase-space area. Conservation of phase space area enclosed by the deformed circle is also confirmed numerically:

$$\frac{э}{2}\int_0^{2\pi} T(\psi)d\psi = \pi\,э.  \tag{7.8}$$

While phase space area occupied by the beam before and after the lens are the same, the effective area, occupied by the beam, increases as a result of the encounter. Let us determine the increase in effective beam emittance as a square of the product of minimum and maximum values of $T$:

$$\frac{э_{eff}}{э} = \sqrt{T_{min}T_{max}}.  \tag{7.9}$$

The values $T_{max}$, $T_{min}$ are determined numerically from Eq. (7.6) (see Fig. 7). Dependence of the emittance growth on the parameter $\upsilon$ is shown in Fig. 8. Qualitatively, this relationship can be approximated by the function:

$$\frac{\unicode{x42D}_{eff}}{\unicode{x42D}} = \sqrt{1+K\upsilon^2} \; , \tag{7.10}$$

where the parameter $K \approx 0.4$. Substitution of Eq. (7.7) into Eq. (7.10) gives the following expression for effective beam emittance growth:

$$\frac{\unicode{x42D}_{eff}}{\unicode{x42D}} = \sqrt{1+K(\frac{C_\alpha R^4}{f \unicode{x42D}})^2} \; . \tag{7.11}$$

Equation (7.11), was tested numerically for a round beam with different particle distributions (see Fig. 9). Simulations were performed with macroparticle code BEAMPATH [14]. As a measure of effective beam emitance, the four-rms beam emittance was used and a two-rms beam size was used as a measure of the beam radius:

$$\unicode{x42D} = 4\sqrt{<x^2><x'^2> - <xx'>^2} \; , \qquad R = 2\sqrt{<x^2>} \; . \tag{7.12}$$

Simulations confirm that dependence, Eq. (7.11), is valid for four-rms beam emittance, although the coefficient $K$ depends on the beam distribution (see Table 1). Generally, the value of $K$ is smaller than that determined above, except for the case of a Gaussian distribution.

## 8. SPACE CHARGE INDUCED BEAM EMITANCE GROWTH IN DRIFT SPACE

Non-linear space charge forces inherent to a non-uniform beam act on the beam as a non-linear lens. Analysis developed in Section 7 can be applied to determine space-charge induced beam emittance growth. Consider the initial stage of beam drift in free space at a certain distance $z$, where radial particle positions are not changed significantly, but the momentum distribution is already affected by the space charge field of the beam, $R=2\sqrt{<r^2>}$. Change in the radial slope of the particle trajectory is given by

$$r' = r'_o + \frac{qzE_b(r)}{mc^2\beta^2\gamma^3} \; . \tag{8.1}$$

Consider a Gaussian beam in drift space. The space charge field of the beam is approximated as:

$$E_b(r) = \frac{I}{2\pi\varepsilon_o\beta c r}[1-\exp(-2\frac{r^2}{R^2})] \approx \frac{I}{\pi\varepsilon_o\beta c}\frac{r}{R^2}(1-\frac{r^2}{R^2}+....) \; . \tag{8.2}$$

Substitution of Eq. (8.2), into Eq. (8.1) results in a transformation, Eq. (6.6), where

$$\frac{1}{f} = -4\frac{I}{I_c(\beta\gamma)^3}\frac{z}{R^2} \; , \qquad C_\alpha = -\frac{1}{R^2} \; . \tag{8.3}$$

Parameter $\upsilon$, Eq. (7.7), which determines the effect of spherical abberation on the beam emittance is therefore

$$\frac{C_\alpha R^4}{f\, э} = \frac{4}{\beta^3 \gamma^3} \frac{I}{I_c} \frac{z}{э} \quad . \tag{8.4}$$

Substitution of Eq. (8.4) into Eq. (7.11) results in the following expression for space charge induced beam emittance growth in free space:

$$\frac{э_{eff}}{э} = \sqrt{1 + \bar{K}(\frac{I}{I_c \beta^3 \gamma^3} \frac{z}{э})^2} \quad . \tag{8.5}$$

The parameter $\bar{K}$ was determined numerically for different distributions (see Fig. 10). Results are summarized in Table 1. As follows from Eq. (8.5), initial emittance growth does not depend on initial beam radius. The same result was obtained in Ref. [15] for a waterbag distribution with $э = 0$.

## SUMMARY

In this work, we have determined matched beam transport conditions for a periodic structure of focusing solenoids in both, emittance-dominated and space-charge-dominated regimes. A closed-form expression for the maximal limited beam current and beam emittance growth in the considered structure were obtained. The developed analysis can be applied to beam matching in a typical solenoid focusing channels.

## ACKNOLEDGEMETS

Author is indebted to Konstantin Batygin for useful discussions and help in preparation of manuscript.

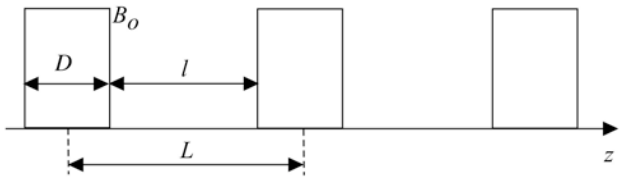

Fig. 1. Periodic structure of focusing solenoids.

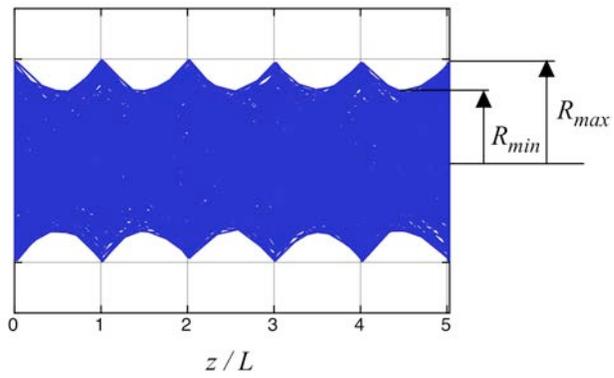

Fig. 2. A matched beam in a periodic focusing structure.

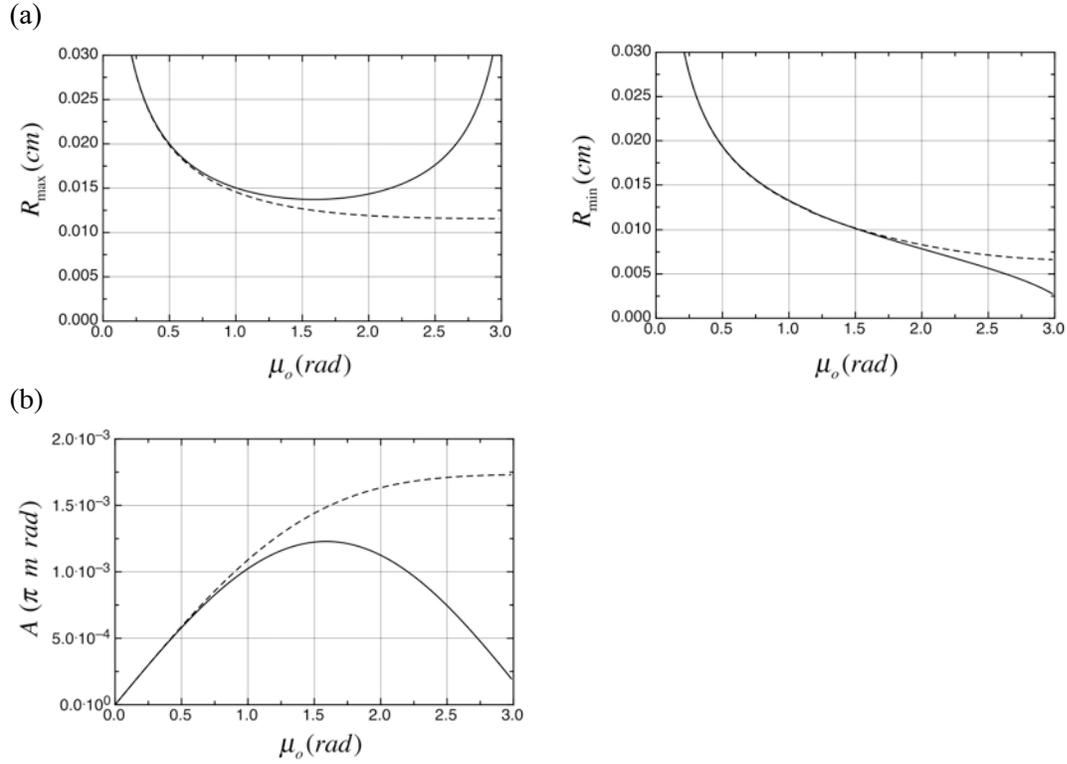

Fig. 3. (a) minimum and maximum beam sizes in a periodic solenoid structure with $D/L = 0.034$: (solid line) solution from matrix analysis, Eqs. (1.6), (1.7), (dotted line) smooth approximation to the beam envelope, Eq. (3.26), (b) acceptance of the channel: (solid line) determined by matrix method, Eq. (1.8), (dotted line) determined from envelope equation, Eq. (3.29).

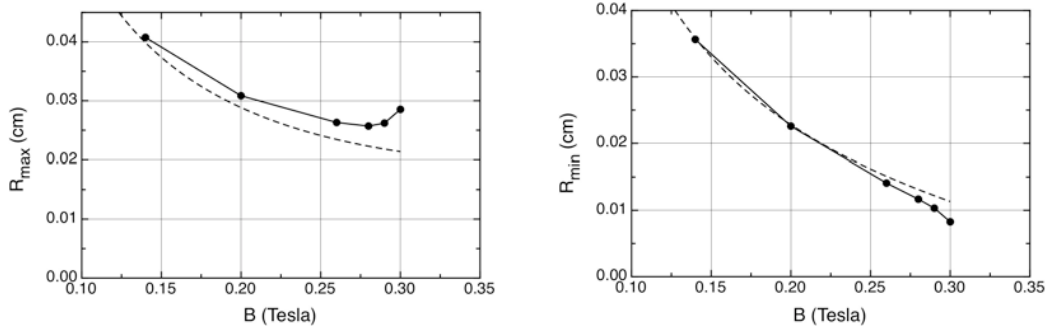

Fig. 4. Minimum and maximum beam sizes in a periodic solenoid structure with $D/L = 0.034$ for a space-charge dominated proton beam with energy $W = 35$ keV, beam current $I = 3.5$ mA and beam emittance $э = 9.262\ \pi\ cm\ mrad$: (solid line) exact solution of envelope equation, Eq. (3.1), (dotted line) smooth approximation to beam envelope, Eq. (3.26).

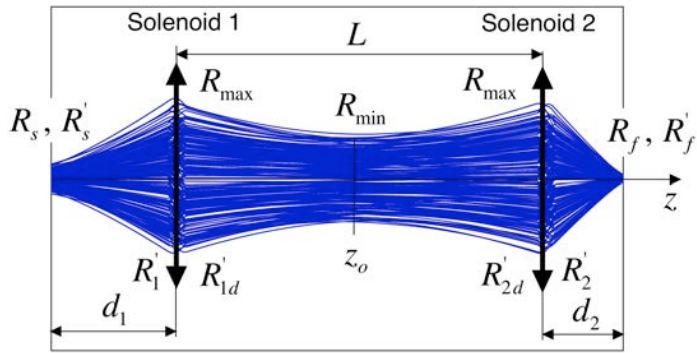

Fig. 5: LEBT with two focusing solenoids.

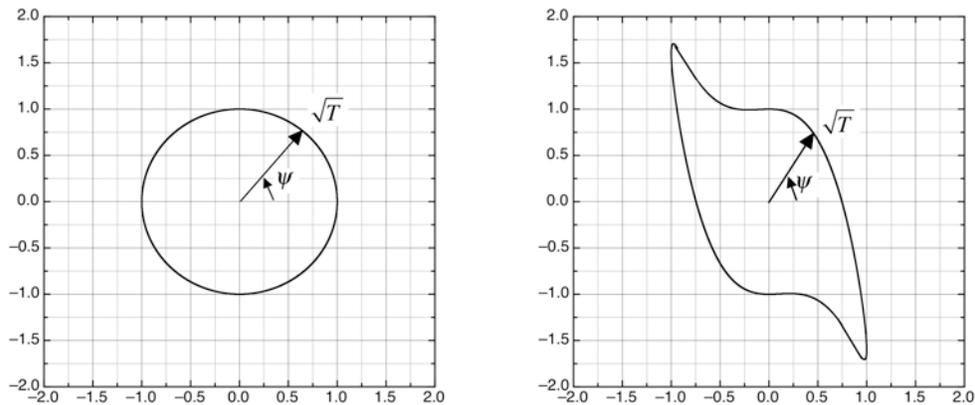

Fig. 6. Distortion of beam emittance due to spherical aberrations, Eq. (7.6): (left) $υ = 0$, (right) $υ = 1.6$.

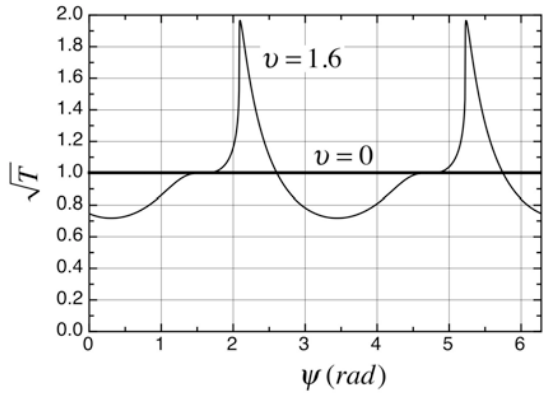

Fig. 7. Action-angle dependence determined by Eq. (7.6).

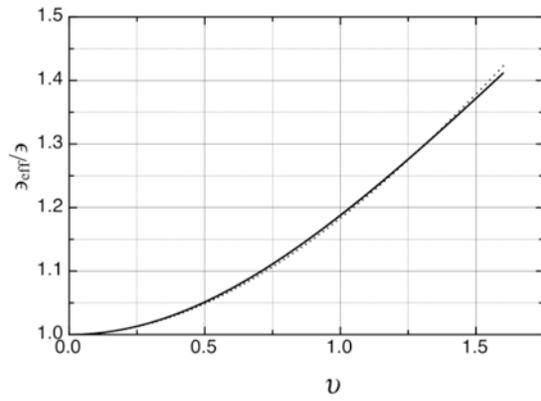

Fig. 8. Effective beam emittance growth due to spherical aberration as a function of the parameter $\upsilon$: (sold line) Eq. (7.9), (dotted line) approximated by Eq. (7.10).

KV

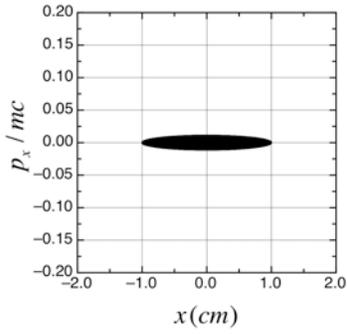 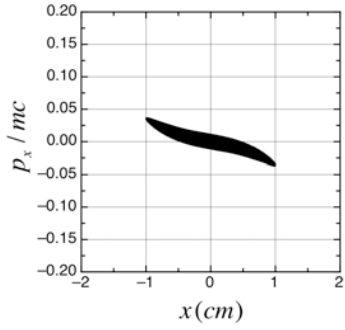

Water Bag

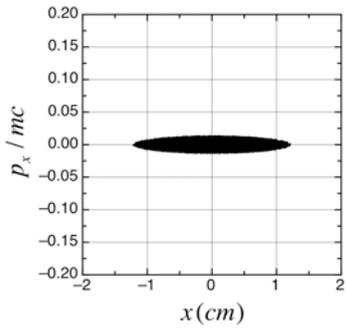 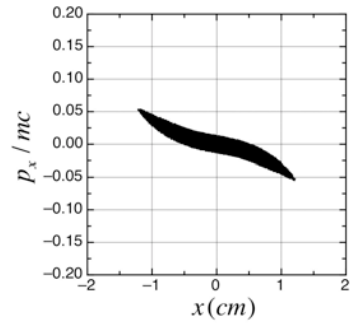

Parabolic

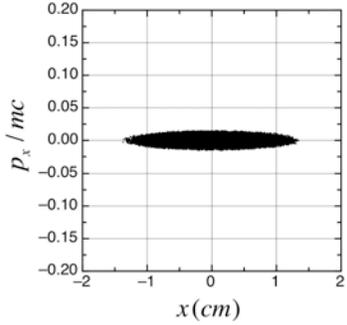 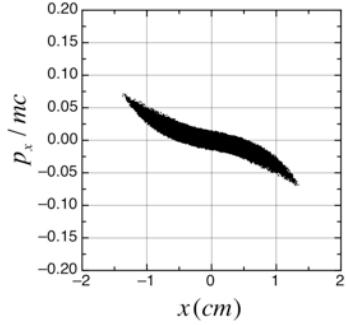

Gaussian

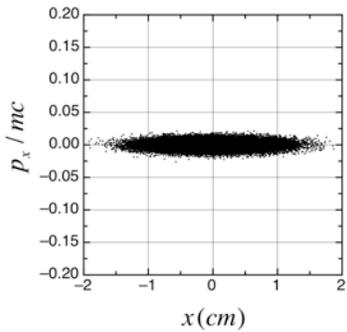 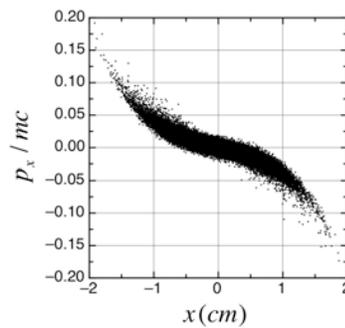

Fig. 9. Effective emittance growth due to spherical aberrations for beams with different initial distributions in a lens with $\upsilon =1.6$, Eq. (7.7).

KV

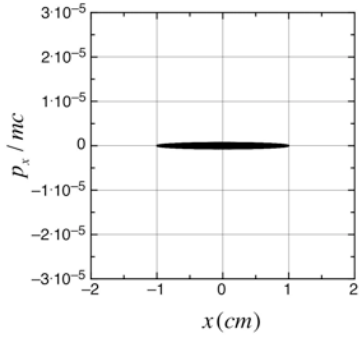
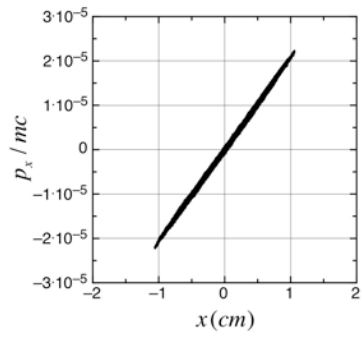

Water Bag

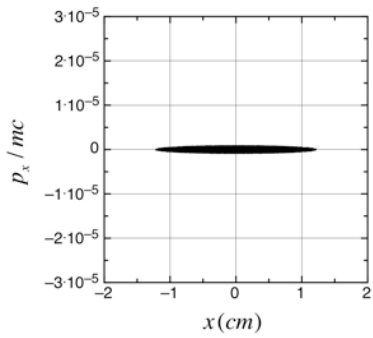
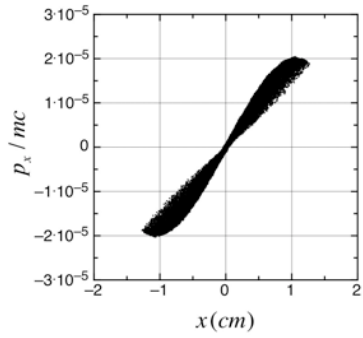

Parabolic

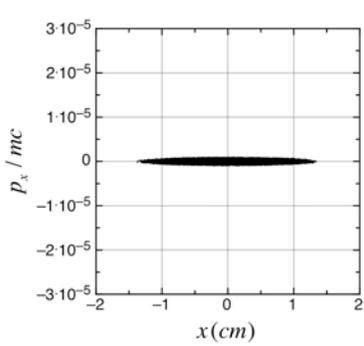
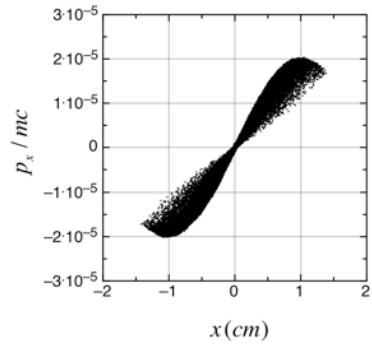

Gaussian

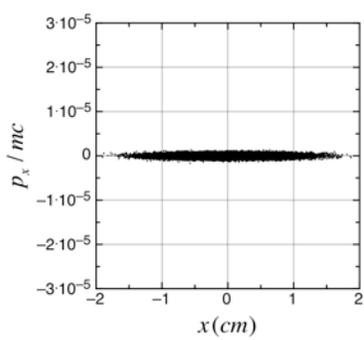
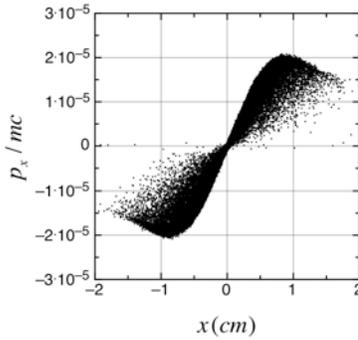

Fig. 10. Emittance growth of a 150 keV proton beam with current $I = 1.11$ mA and beam emittance $э = 0.039 \, \pi$ cm mrad in drift space of length $z = 100$ cm.